\documentclass{article}
\usepackage{spconf,amsmath,graphicx}

\usepackage{balance}
\usepackage{xcolor}
\usepackage{multirow}
\usepackage{adjustbox}
\usepackage{soul}
\usepackage{setspace}
\usepackage{eucal}
\usepackage{booktabs}
\usepackage{pifont}
\usepackage{hyperref}
\usepackage{amssymb}
\usepackage{pgfplots}
\usepackage{pgfplotstable}
\pgfplotsset{compat=1.7}
\usepackage{tikz}

\usepackage{fancyhdr}
\usepackage{lipsum}
\fancypagestyle{mystyle}{%
    \fancyhead[R]{To appear in ICASSP 2023}
}%

\title{Heterogeneous Graph Learning for Acoustic Event Classification}
\name{Amir Shirian$^1$, Mona Ahmadian$^2$, Krishna Somandepalli$^3$, Tanaya Guha$^{4}$}
\address{
  $^1$University of Warwick, UK, 
  $^2$University of Surrey, UK,
  $^3$Google Research, USA,\\
  $^4$University of Glasgow, UK}

\begin{document}
\ninept
\maketitle
\thispagestyle{mystyle}
\begin{abstract}
Heterogeneous graphs provide a compact, efficient, and scalable way to model data involving multiple disparate modalities. This makes modeling audiovisual data using heterogeneous graphs an attractive option. However, graph structure does not appear naturally in audiovisual data. Graphs for audiovisual data are constructed manually which is both difficult and sub-optimal. In this work, we address this problem by (i) proposing a parametric graph construction strategy for the intra-modal edges, and (ii) \emph{learning} the crossmodal edges. To this end, we develop a new model, \emph{heterogeneous graph crossmodal network} (HGCN) that learns the crossmodal edges. Our proposed model can adapt to various spatial and temporal scales owing to its parametric construction, while the learnable crossmodal edges effectively connect the relevant nodes across modalities. Experiments on a large benchmark dataset (\textit{AudioSet}) show that our model is state-of-the-art ($0.53$ mean average precision), outperforming transformer-based models and other graph-based models. Our code is available at \mbox{\href{https://github.com/AmirSh15/Cross_modality_graph}{\texttt{github.com/AmirSh15/Cross\_modality\_graph}}}
\end{abstract}


\section{Introduction}
\label{sec:intro}
Visual information is known to augment and complement human perception and cognition of audio events \cite{mcgurk1976hearing, atilgan2018integration}. Therefore, learning audiovisual representations is critical to improve performance of various audio classification tasks. For example, consider the task of identifying an acoustic event, where a motorbike is moving away from the microphone. The revving sound of the bike fades as it moves away. While an audio-only model may not be able to identify the fading sound as `motorbike', adding a visual clip of the motorbike moving does help. 

A majority of existing works on learning audiovisual representations relies on models that are originally developed to address computer vision tasks \cite{alayrac2020self,MaZMS21}. They usually augment two `views' of a given audiovisual sample, which are then fed to a shared `backbone' model trained using a suitable optimization function, such as contrastive loss \cite{saeed2021contrastive, ma2021contrastive}, distillation loss \cite{chen2021distilling} or information maximization \cite{shukla2020learning, zbontar2021barlow}. The above models, however, do not fully capture the temporal relationship between the two modalities. Moreover, the vision-inspired data augmentation techniques are often unsuitable for multimodal data \cite{falcon2022feature}.

Heterogeneous graphs provide a compact, efficient, and scalable way to model data involving multiple disparate modalities (e.g., image and text) and their relationships \cite{wang2021dualgnn, qian2021dual}. Hetergeneous multimodal graphs have been successfully used to address tasks such as visual question answering \cite{saqur2020multimodal}, multimodal sentiment analysis \cite{yang2021multimodal}, and cross-modal retrieval \cite{qian2021dual}. These recent works have established that a multimodal graph approach promotes closer coupling between various events across the modalities resulting in a significant performance gain \cite{saqur2020multimodal,yang2021multimodal}. 

In our past work \cite{shirian2022visually}, we noted that hetereogenous audiovisual graphs can effectively capture the relationship within and across audio and visual modalities, which can outperform other multimodal learning approaches. However, the success of this approach, to a large extent, relies on constructing the `right' graph. Since the graph structure is not naturally known here, it is difficult (and still sub-optimal) to construct the `right' graph. The current paper addresses this important issue of multimodal graph construction by learning the graph structure across the modalities in the context of acoustic event classification. The idea of crossmodal learning on graphs has been successfully used to capture semantics within a modality and semantic interactions between them in applications involving vision and language \cite{wang2021wasserstein,song2021spatial}. Motivated by this success we propose a novel graph-based approach to learning crossmodal interactions between audio and video for acoustic event classification.
%
%
\begin{figure*}
    \centering
    \includegraphics[width=1.0\linewidth,height=5cm]{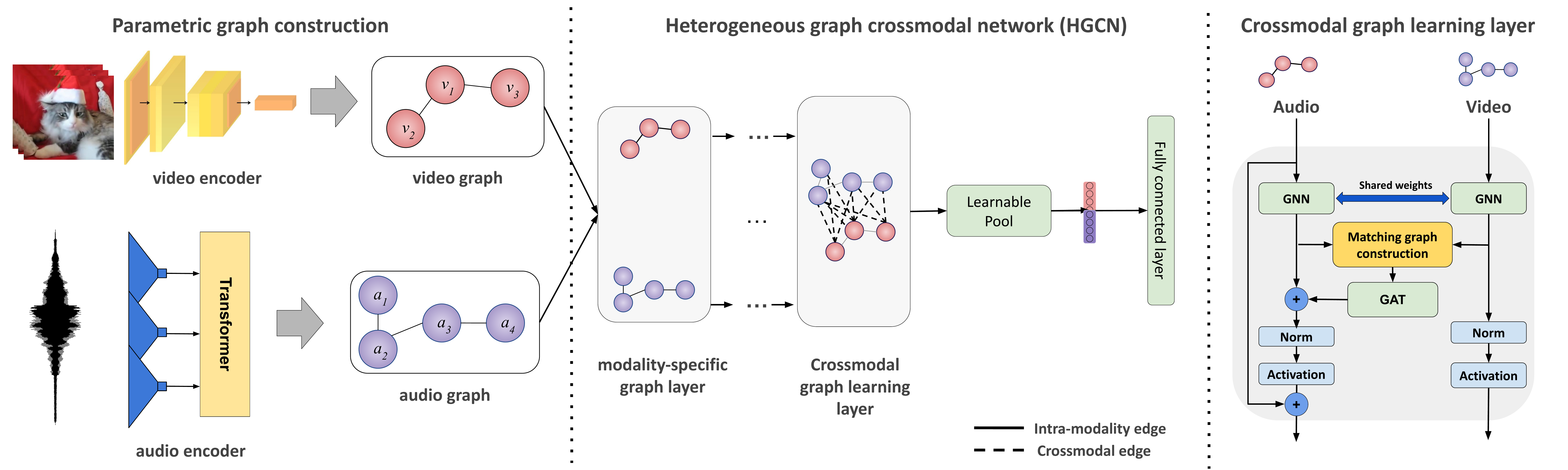}
    \caption{\textbf{Our end-to-end graph-based approach}: [Left] We split the audio/video input into non-overlapping segments, and construct a subgraph for each modality. (Centre) These subgraphs are combined into a single graph that has only intra-modal edges, and processed through a modality-specific graph layer. For both audio and video modalities independently, modality-specific graph convolution layers are utilised to extract the embedding for each node. Next, the crossmodal edges are learned through the proposed heterogeneous crossmodal layer. Learnable pooling modules are used to capture the overall graph representation. Our \textit{crossmodal graph learning layer} has two independent audio and video branches, a shared graph neural network, and a matching graph construction module following an attention layer connecting video nodes to audio nodes considering the inter-modality edges constructed by the matching graph. 
    }
   \vspace{-3mm}
    \label{fig:overview}
\end{figure*}
\par In this paper, we propose an end-to-end graph approach to acoustic event classification that learns audio representation utilizing heterogeneous graphs. The key contribution is to parameterize the process of graph construction and learning the crossmodal edges along with the task. First we construct modality-specific subgraphs (controlled by two parameters), which are fed to our \emph{heterogeneous graph crossmodal network} (HGCN). The key feature of HGCN is a \emph{crossmodal graph learning layer} that learns graph edges across modalities jointly with the classification task. HGCN also has a modality-specific graph layer that can perform modality-specific graph processing. Our model, HGCN, thus allows for both independent processing of each modality and fusing information in the crossmodal layer. The idea presented in this paper is significantly different from previous graph-based approaches used for representation learning \cite{shirian2022visually,shirian2021compact} as it avoids manually connecting nodes and makes end-to-end learning possible. In summary, our contributions are as follows:
\begin{itemize}
    \item We develop an end-to-end deep graph approach to audio representation learning from heterogeneous audiovisual graphs.
    \item We propose a parametric graph construction strategy and the HGCN model with a novel crossmodal graph learning layer. Our model can capture modality-specific information as well as complementary information between the two modalities.
    \item We demonstrate state-of-the-art performance of our model for the task of acoustic event classification on a large benchmark dataset called the \textit{AudioSet}.
\end{itemize}
%
%

\section{Proposed Approach}
\label{sec:proposed}
In this section, we describe our end-to-end graph approach in detail. It has two components: (i) \emph{subgraph construction} and (ii) the \emph{HGCN model}. In the subgraph construction stage, we construct individual graphs for each modality. This process is controlled by two parameters. These subgraphs are combined into a single graph that has only intra-modal edges. This graph is fed to our proposed model, HGCN, which has a modality-specific graph processing layer and a crossmodal graph learning layer. HGCN learns of the crossmodal graph edges jointly with the classification task.

\vspace{2mm}
\par\noindent\textbf{Definition:} We define a \emph{heterogeneous graph} $\mathcal{G}$ to be composed of an audio subgraph and a video subgraph. This can be represented as $\mathcal{G} = (\mathcal{V},\mathcal{E})$, where $\mathcal{V}=\{\mathcal{V}_a, \mathcal{V}_v\}$ is the set of audio nodes $\mathcal{V}_a$ and video nodes $\mathcal{V}_v$; $\mathcal{E}=\{\mathcal{E}_{aa}, \mathcal{E}_{vv}, \mathcal{E}_{av}\}$ is three sets of edges: audio-audio, video-video and audio-video.
\vspace{-3mm}
\subsection{Audio and video subgraph construction}
\begin{figure}[t]
    \centering
    \includegraphics[width=1.0\linewidth, trim=1.1cm 1.1cm 1.2cm 0.3cm, clip=true]{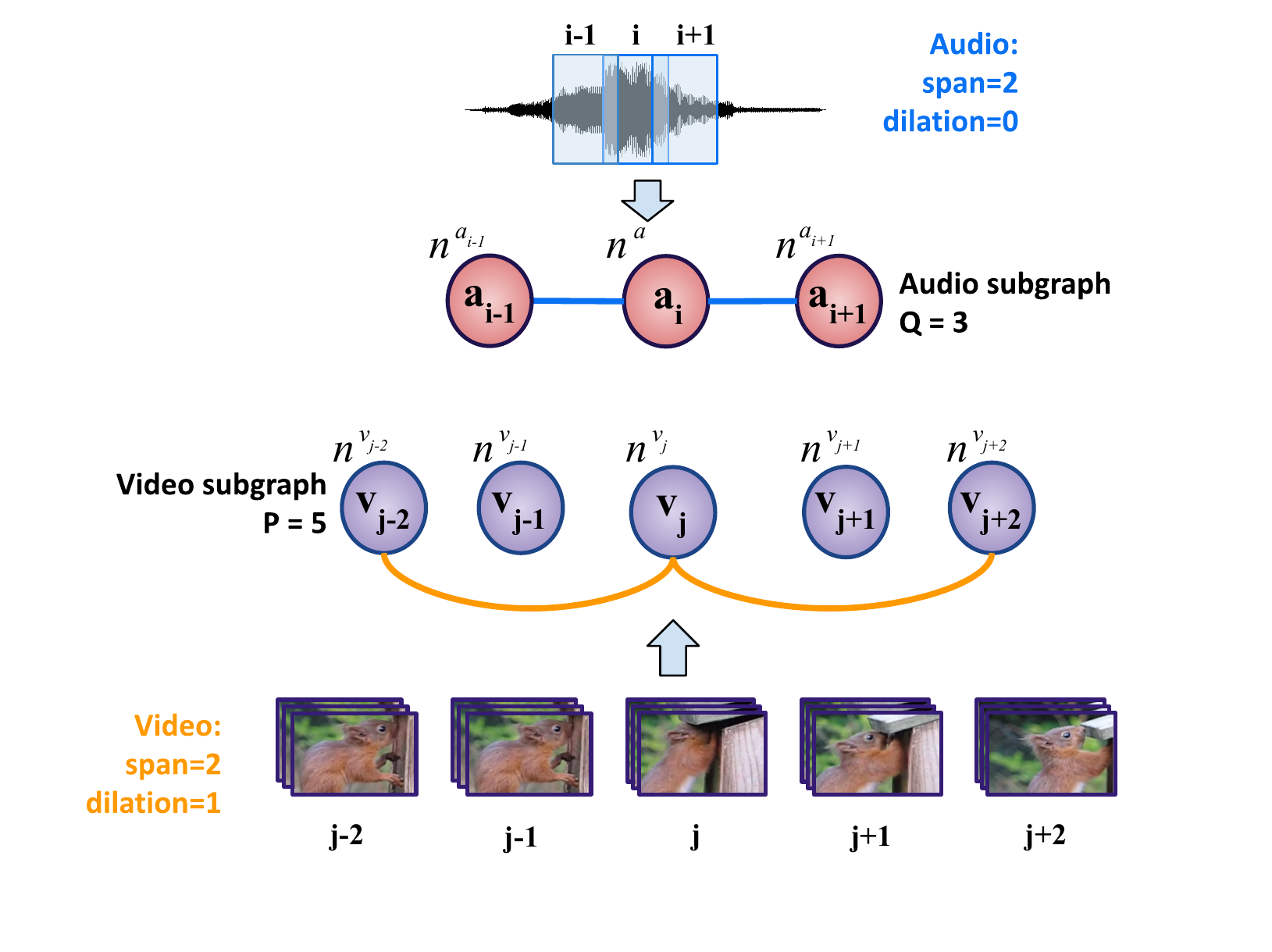}
    \caption{Audio and video subgraph construction process: For simplicity, the edges are only shown for one node per modality, $v_i$ for audio and $v_j$ for video. Crossmodal edge are \emph{learned} in the crossmodal graph learning layer within the proposed HGCN model. 
    }
    \vspace{-5mm}
    \label{fig:graph_const}
\end{figure}

%
The first step is to construct the audio and video subgraphs. Given an audiovisual input, we divide the audio and the video into $Q$ and $P$ segments (see  Fig.\ref{fig:graph_const}). Each segment corresponds to a node. Therefore, the audio subgraph has $Q$ nodes and the video subgraph has $P$ nodes. These segments are used for feature extraction, so as to have a single feature vector per node. Therefore, $\mathcal{G}$ has audio node set $\mathcal{V}_a = \{a_i\}_{i=1}^Q$ and video node set $\mathcal{V}_v = \{v_i\}_{i=1}^P$, with edge sets $\mathcal{E}=\{\mathcal{E}_{aa},\mathcal{E}_{vv}\}$. At this stage we work with intra-modal edges only, and hence $\mathcal{E}_{av}$ is empty. Each node $v_i\in\mathcal{V}_v$ corresponds to a video segment and its associated with a feature vector $\mathbf{n}_i^v$ extracted by a video encoder. Similarly, every audio node $a_i\in\mathcal{V}_a$ is associated with a feature vector $\mathbf{n}_i^a$ obtained from an audio encoder. 

We propose to add intra-modality edges using two parameters for each intramodal edge type, $\mathcal{E}_{vv},\mathcal{E}_{aa}$. These parameters are (i) \textit{span across time} and (ii) \textit{dilation}. For a given node, \emph{span across time}; \textit{span} for brevity, denotes the number of nodes it connects with in the temporal direction, whereas \emph{dilation} denotes the skip between connections. For example, in Fig.\ref{fig:graph_const}, dilation for audio is 0 as consecutive nodes are connected; while for video it is 1, since we skip one node between connections. We have $4$ hyperparameters in total, $2$ for each node type. This provides more control on the graph construction process as each modality can be modeled with their own parameters.
\subsection{Learning crossmodal edges}
The next step is to learn the edges between audio and video nodes i.e. the edges in $\mathcal{E}_{av}$. To this end, we propose a new model, HGCN, which has \emph{two} types of graph processing layers: (i) \textit{modality-specific} graph layer and (ii) \emph{crossmodal graph learning} layer (see Fig.\ref{fig:overview}). 

\vspace{1mm}
\noindent\textbf{Modality-specific graph layer:} The key idea in the majority of graph neural network (GNNs) is to aggregate features from a node's neighbours and update that node's feature, $\mathbf{H}$ accordingly: 
\begin{equation}\label{eq:gnn}
\begin{aligned}
  \mathbf{H}_{l+1} = \sigma\big(\mathbf{A}\mathbf{H}_l\mathbf{W}_l\big)
\end{aligned}
\end{equation}
where $\mathbf{W}_l$ is the weight matrix for the $l^{th}$ layer, $\mathbf{A}$ is the adjacency matrix, $\sigma$ is a non-linear activation function, such as ReLU, and $l\in \{0, \cdots L\}$. Since we have a heterogeneous graph with different nodes and edge types, this approach is not directly applicable. 

Previous studies have utilised meta-paths for processing heterogeneous graphs \cite{fu2020magnn,hu2020heterogeneous}, which was shown to be inadequate in capturing the information provided by disparate nodes and edge types \cite{lv2021we}. To address this, we use separate GNNs for processing different edge types. This is done in the modality-specific graph layer of our HGCN. As the name suggests, the modality-specific layer processes each modality independently, considering only the intra-modality edges. Therefore the node attributes are updated as follows:
\begin{equation}\label{eq:het_graph}
    \begin{aligned}
        \mathbf{n}_{l+1}^a &= \text{GNN}_{\theta_1}\big( \mathbf{n}_l^a, \mathbf{A}_a \big) \\
        \mathbf{n}_{l+1}^v &= \text{GNN}_{\theta_2}\big( \mathbf{n}_l^v,\mathbf{A}_v \big)
    \end{aligned}
\end{equation}
where $\mathbf{n}_{l}^a$ and $\mathbf{n}_{l}^v$ are audio and video node features in layer \textit{l}, and GNN can be any graph-based neural network such as GCN \cite{kipf2017semi}, GraphSage \cite{hamilton2017inductive}, or GAT \cite{velivckovic2017graph}. $\mathbf{A}_a$ and $\mathbf{A}_v$ are the adjacency matrices pertaining to the audio and video subgraphs.

\vspace{1mm}
\noindent\textbf{Crossmodal graph learning layer:} This layer learns the crossmodal edges jointly with the classification objective. The layer first establishes a shared space between the audio and visual modalities through a shared GNN using the embedding from the previous multimodal graph layers. Then, the layer constructs a \emph{matching graph} between the nodes of two modalities. Finally, a fusion flow carries audio-related information from the video nodes to the audio nodes for the inter-modality edges ($\mathcal{E}_{av}$) and the corresponding adjacency matrix $\mathbf{A}_{av}$ is constructed.
\begin{equation}\label{eq:het_crossmodal}
    \begin{aligned}
        \mathbf{n}_{l+1}^a &= \text{GNN}_{\phi_1}\big( \mathbf{n}_l^a, \mathbf{A}_a \big) + \text{GNN}_{\phi_2}\big( \mathbf{n}_l^v, \mathbf{A}_{av} \big)\\
        \mathbf{n}_{l+1}^v &= \text{GNN}_{\phi_1}\big( \mathbf{n}_l^v,\mathbf{A}_v \big)
    \end{aligned}
\end{equation}
Note that we used a shared GNN in \eqref{eq:het_crossmodal}. The video nodes are only updated using video nodes from the previous layer, but the audio nodes, audio being the primary source of information for our task, use information from both audio and video nodes.

The key feature of this layer is a \emph{matching graph construction} module that connects neighbouring nodes from different modalities. These neighboring nodes are computed using k-nearest neighbour:
\begin{equation} \label{eq:knn}
\begin{aligned}
    \mathcal{E}_{av} = \underset{d\big(\mathbf{n}_{l}^{a_i},\mathbf{n}_{l}^{v_j}\big)}{\text{Top k }}\Big\{(i,j) | i \in \{1, \cdots, Q \}, j \in \{1, \cdots, P \} \Big\} 
\end{aligned}
\end{equation}
where $d$ is the cosine or L2 distance. The advantages of this layer are: (i) it reduces redundant connections, and (ii) detects relevant nodes between modalities instead of densely attending to all nodes.

Finally, we seek a graph-level representation $\mathbf{h}_G\in\mathbb{R}^d$ as the output of HGCN. This is obtained by pooling the node-level representations $\mathbf{n}_L^a$, $\mathbf{n}_L^v$ at the $L$-th layer before passing them to the classification layer. Common pooling functions (\texttt{mean}, \texttt{max} and \texttt{sum} pooling) treat adjacent nodes with equal importance, which may not be optimal. Thus we \textit{learn} a pooling function $\Psi$ following a recent work \cite{shirian2021dynamic} that combines the node embeddings from the $K$-th layer to produce an embedding for the entire graph. The pooling layer is defined as follows: 
\begin{equation} \label{eq:graph_emb}
\begin{aligned}
    \mathbf{h}_G = \Big[\Psi_a(\mathbf{n}_L^a) \,|\,\Psi_v(\mathbf{n}_L^v) \Big]=\mathbf{p}^a\mathbf{n}_L^a  + \mathbf{p}^v \mathbf{n}_L^v 
\end{aligned}
\end{equation}
where $\mathbf{p}^a$ and $\mathbf{p}^v$ are learnable weights. The overall heterogeneous graph network is trained with the cross-entropy loss.

\section{Experiments}
\label{sec:Exp}
\subsection{Dataset}
\vspace{-1mm}
We use a large scale weakly labelled dataset called the \textbf{AudioSet} \cite{gemmeke2017audio}, which contains audio segments from YouTube videos. We work with 33 classes from the balanced set that have high rater confidence score ($\{0.7, 1.0\}$). This yields a training set of 82,410 auviovisual clips. For a fair comparison with baseline methods, we used the original evaluation set, which has 85,487 test clips.
\subsection{Feature encoder}
\vspace{-1mm}
\noindent \textbf{Audio encoder:} To extract audio node features, each audio clip is divided into $320$ ms non-overlapping segments. Wav2vec2 \cite{baevski2020wav2vec} (pretrained on LibriSpeech) feature extractor is used as an audio encoder. For each segment, temporal convolution layers have been used to extract features for the receptive field of 25 ms and stride of 20 ms.
We use the 512-dimensional features extracted by the wave2vec2 network for each segment.

\noindent \textbf{Video encoder:} Each video is sampled at 20 fps and resized to $112\times112$, and then segmented into non-overlapping 1s chunks to extract the video node features. As a video encoder, we used R(2+1)D \cite{tran2018closer} pretrained on Kinetics-400. A 512-dimensional feature is then obtained by feeding each segment into the network. Note that our method is not tied to these networks and can work with any generic encoder for both audio and video.

\subsection{Implementation details}
\vspace{-1mm}
Each video clip is transformed to a heterogeneous graph with $P=10$ audio nodes and $Q=30$ video nodes. Each audio node corresponds to a 320ms-long audio segment and each video node corresponds to a 1s-long video segment. We use  $3$ nearest neighbors while learning the crossmodal matching graph. To understand robustness of this step, we repeat our experiments $10$ times with different seeds and report both mAP (mean average precision) and ROC-AUC (area under the ROC curve) values. 

The network weights are initialized following the Xavier initialization. We used SGD optimizer with a learning rate of $0.005$ for the graph model, $5\times10^{-4}$ for audio and video encoders, and $1000$ warm-up iterations for all experiments. The graph construction hyperparameters are explored heuristically and set to \textit{span audio} = 3, \textit{dilation audio} = 1, \textit{span video} = 1, and \textit{dilation video} = 1 for all experiments. For the GNN, we select GraphSage \cite{hamilton2017inductive} for the shared GNN in the crossmodal layer, and modality-specific layers, and a Graph Attention Network (GAT) \cite{velivckovic2017graph} for fusing information in the crossmodal graph learning layer. We have three modality-specific graph layers and one crossmodal graph learning layer (see Fig. \ref{fig:overview}), each with a hidden size of $512$. We use Pytorch on an NVIDIA RTX-2080Ti GPU. 

\begin{table}[t]
\centering
\caption{Acoustic event classification results on \textbf{AudioSet}. 
}
\vspace{2mm}
\resizebox{1.0\linewidth}{!}{
\renewcommand*{\arraystretch}{1.0}
  \begin{tabular}{l|c|c|c}\cline{2-4}
        \toprule
        \bf Model & \bf mAP & \bf ROC-AUC & \bf Params          \\ 
        \midrule
        Ours (audio only)   & $0.46\pm0.06$ & $0.89\pm0.03$ & $40.2$M\\   
        Ours (video only) & $0.38\pm0.03$ & $0.84\pm0.02$ & $39.8$M\\
        Ours (fixed encoders) & $0.50 \pm0.02$ & $0.90\pm0.02$ & $42.4$M\\
        \textbf{Ours} (end-to-end) & $\mathbf{0.53} \pm0.01$ & $\mathbf{0.94}\pm0.01$ & $42.4$M\\
        \midrule
        \multicolumn{4}{c}{\emph{Baselines}}  \\
        \midrule
        Wav2vec2 audio only & $0.42 \pm0.02$ & $0.88\pm0.00$ & $94.4$M\\
        R(2+1)D video only & $0.36 \pm0.00$ & $0.81\pm0.00$ & $33.4$M\\
        \midrule
        \multicolumn{4}{c}{\emph{State-of-the-art}}  \\
        \midrule
        DaiNet \cite{dai2017very} & $0.25\pm0.07$ & - & $1.8$M \\
        Spectrogram-VGG   & $0.26\pm0.01$ & - & $6$M\\
        VATT \cite{akbari2021vatt} & $0.39\pm0.02$ & - & $87$M  \\
        SSL graph \cite{shirian2022self}   & $0.42\pm0.02$  & - & $218$K  \\
        Wave-Logmel \cite{kong2020panns} & $0.43\pm0.04$  & - & $81$M  \\
       AST \cite{gong2021ast} & $0.44\pm0.00$ & -  & $88$M  \\
       VAED \cite{shirian2022visually} & $0.50\pm0.01$ & $0.93\pm0.00$  & $2.1$M  \\
        \bottomrule
        \end{tabular}
        }
        \label{tab:Auco_Eve_Clas}
        \vspace{-3.5mm}
\end{table}

\subsection{Results and analysis}
\noindent \textbf{Baselines:} Table \ref{tab:Auco_Eve_Clas} compares our model with two competitive and relevant self-supervised models: wav2vec2 \cite{baevski2020wav2vec} and R(2+1)D \cite{tran2018closer} to investigate the superiority of our end-to-end approach.

\vspace{2mm}
\noindent \textbf{State-of-the-art:} We also compare our method with a number of strong supervised and self-supervised state-of-the-art models. 
The DaiNet \cite{dai2017very} is a 1D convolution-based network which operates on raw audio waveform. 
The Spectrogram-VGG model is the same as the configuration A in \cite{simonyan2014very} with only one change: the final layer is a softmax with 33 units. The feature for each audio input to the VGG model is a log-mel spectrogram of dimensions 96$\times$64 computed by averaging across non-overlapping segments of length 960ms. 
The VATT \cite{akbari2021vatt} is a self-supervised multimodal transformer with a modality-agnostic, single-backbone Transformer and sharing weights between audio and video modality.
We also compared our method with recent graph-based works \cite{shirian2022self,shirian2022visually}.
The wave-Logmel \cite{kong2020panns} is a supervised CNN model which takes waveform and log mel spectrogram at the same time as input.
The AST \cite{gong2021ast} is a self-supervised transformer model which is trained by masking the input spectrogram.
All methods' hyper-parameters are set to the values published in the original papers. Note that we do not utilise any data augmentation, despite the fact that other methods used powerful data augmentations. Additionally, all of the baselines have been retrained using the same classes as our model.

\vspace{2mm}
\noindent \textbf{Results:} Table \ref{tab:Auco_Eve_Clas} compares the performance of our model with the baselines and various state-of-the-art models in terms of mAP and ROC-AUC with their standard deviation. Our model, HGCN, outperforms all baselines and state-of-the-art models. HGCN outperforms the VAED model \cite{shirian2022visually}, state-of-the-art on AudioSet, by more than $3\%$ in mAP. Overall, HGCN, achieves a superior mAP score demonstrating the effectiveness of our graph construction and cross-modal fusion strategies. Furthermore, our model achieves the highest ROC-AUC score ($0.94$) indicating more trustworthy predictions at various thresholds. Also note that HGCN has significantly fewer learnable parameters compared with the recent transformer-based architectures, i.e., VATT \cite{akbari2021vatt} and AST \cite{gong2021ast}.
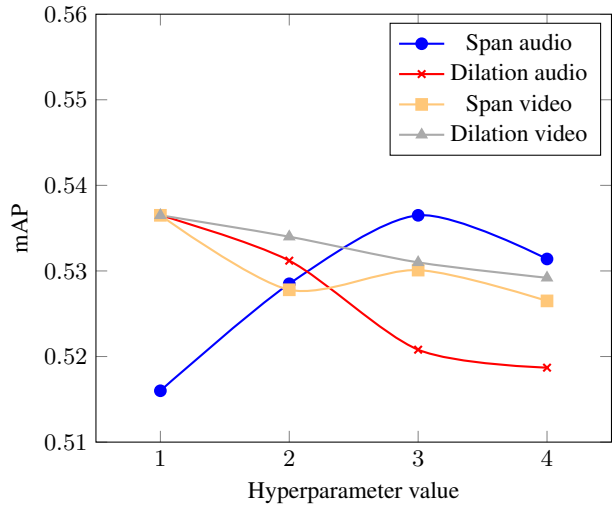
\begin{figure}[!ht]
    \begin{minipage}[t]{\linewidth}
    \centering
    \begin{tikzpicture}
        \begin{axis}[
            xlabel=Hyperparameter value,
            ylabel=mAP,
            xmin=0.5, xmax=4.5,
            ymin=0.51, ymax=0.56,
            xtick={1,2,3,4},
            every axis plot/.append style={thick}
            ]
        \addplot[smooth,mark=*,blue] plot coordinates {
            (1,0.516)
            (2,0.5285)
            (3,0.5365)
            (4,0.5314)
        };
        \addlegendentry{Span audio}

        \addplot[smooth,color=red,mark=x]
        plot coordinates {
            (1,0.5365)
            (2,0.5312)
            (3,0.5208)
            (4,0.5187)
        };
        \addlegendentry{Dilation audio}

        \addplot[smooth,color={rgb:orange,1;yellow,2;pink,5},mark=square*]
        plot coordinates {
            (1,0.5365)
            (2,0.5278)
            (3,0.5301)
            (4,0.5265)
        };
        \addlegendentry{Span video}

        \addplot[smooth,color={rgb:black,1;white,2},mark=triangle*]
        plot coordinates {
            (1,0.5365)
            (2,0.534)
            (3,0.531)
            (4,0.5292)
        };
        \addlegendentry{Dilation video}
        \end{axis}
    \end{tikzpicture}
    \end{minipage}
    \caption{Effect of using different graph construction hyperparameters (span and dilation) on model performance.}
    \label{fig:graph_size}
\end{figure}

\vspace{1mm}
\noindent \textbf{Ablation experiments:} We conduct in-depth ablation experiments to examine the contribution of each components in our model. Table~\ref{tab:ablation} presents the ablation results in terms of mAP. We observe that each component brings improvement. The addition of the video nodes boosts performance by roughly $5\%$, and when combined with our novel crossmodal graph learningn layer the performance rises by another $4\%$. The learnable pooling  layer improves the mAP score by 1 more percent. 
The ablation results show that each of the proposed components in our architecture is important, and contributes positively towards the overall model performance.
\begin{table}[tb]
\centering
\caption{Ablation experiments showing the contribution of each component in our model.}
\vspace{1mm}
\resizebox{\linewidth}{!}{
\label{tab:ablation}
\small
\renewcommand*{\arraystretch}{1.2}
\begin{tabular}{cccc c}
\hline
{\bf Audio} & {\bf Video } & \bf  Crossmodal  & \bf Learnable pool & {\bf mAP}  \\ 
\hline \hline
\checkmark & - & -     &  - & $0.43$          \\
- & \checkmark & -     &  - & $0.36$          \\
\checkmark & \checkmark & -  &  - & $0.48$        \\
\checkmark & \checkmark & \checkmark  &  - & $0.52$        \\
 \checkmark & \checkmark & \checkmark    & \checkmark &  $\mathbf{0.53}$\\
\hline
\end{tabular}
}
\vspace{-4mm}
\end{table}
%

\vspace{2mm}
\noindent \textbf{Graph construction parameters:}
Fig.~\ref{fig:graph_size}(b) investigates the effect of the graph construction hyperparameters (span and dilation). For audio dilation, audio span, and video dilation performance drops as we increase their values. For audio span, performance improves up to $3$ and then start falling. This explains our choice of these hyperparameters.

\section{Conclusion}
\label{sec:conclusions}
We introduced an end-to-end graph-based approach to audio representation learning with application to acoustic event classification. This involves a parametric modality-specific subgraph construction process, and the \emph{HGCN} model that allows learning crossmodal edges through its crossmodal graph learning layer. Thus we can effectively capture spatial and temporal relationships between audio and visual modalities explicitly. Our model can easily adapt to different temporal scales of events through the span and dilation hyperparameters. Our model is state-of-the-art for acoustic event classification producing highest mAP and ROC-AUC on the AudioSet dataset. Our model relies on separate audio and video encoders, which gives the flexibility of choosing a suitable encoder depending on the application. Currently, we only used one crossmodal layer at the end of the modality-specific GNN layers. More crossmodal learning layers may be useful, and each layer may capture different information shared across modalities. Future work could also be directed towards make the subgraph construction hyperparameters learnable.

\bibliographystyle{IEEEtran}

\bibliography{biblo}

\begin{thebibliography}{10}
\providecommand{\url}[1]{#1}
\csname url@samestyle\endcsname
\providecommand{\newblock}{\relax}
\providecommand{\bibinfo}[2]{#2}
\providecommand{\BIBentrySTDinterwordspacing}{\spaceskip=0pt\relax}
\providecommand{\BIBentryALTinterwordstretchfactor}{4}
\providecommand{\BIBentryALTinterwordspacing}{\spaceskip=\fontdimen2\font plus
\BIBentryALTinterwordstretchfactor\fontdimen3\font minus
  \fontdimen4\font\relax}
\providecommand{\BIBforeignlanguage}[2]{{%
\expandafter\ifx\csname l@#1\endcsname\relax
\typeout{** WARNING: IEEEtran.bst: No hyphenation pattern has been}%
\typeout{** loaded for the language `#1'. Using the pattern for}%
\typeout{** the default language instead.}%
\else
\language=\csname l@#1\endcsname
\fi
#2}}
\providecommand{\BIBdecl}{\relax}
\BIBdecl

\bibitem{mcgurk1976hearing}
H.~McGurk and J.~MacDonald, ``Hearing lips and seeing voices,'' \emph{Nature},
  vol. 264, no. 5588, pp. 746--748, 1976.

\bibitem{atilgan2018integration}
H.~Atilgan, S.~M. Town, K.~C. Wood, G.~P. Jones, R.~K. Maddox, A.~K. Lee, and
  J.~K. Bizley, ``Integration of visual information in auditory cortex promotes
  auditory scene analysis through multisensory binding,'' \emph{Neuron},
  vol.~97, no.~3, pp. 640--655, 2018.

\bibitem{alayrac2020self}
J.-B. Alayrac, A.~Recasens, R.~Schneider, R.~Arandjelovi{\'c}, J.~Ramapuram,
  J.~De~Fauw, L.~Smaira, S.~Dieleman, and A.~Zisserman, ``Self-supervised
  multimodal versatile networks,'' \emph{NeurIPS}, vol.~33, pp. 25--37, 2020.

\bibitem{MaZMS21}
S.~Ma, Z.~Zeng, D.~J. McDuff, and Y.~Song, ``Active contrastive learning of
  audio-visual video representations,'' in \emph{{ICLR}}, 2021.

\bibitem{saeed2021contrastive}
A.~Saeed, D.~Grangier, and N.~Zeghidour, ``Contrastive learning of
  general-purpose audio representations,'' in \emph{ICASSP}.\hskip 1em plus
  0.5em minus 0.4em\relax IEEE, 2021, pp. 3875--3879.

\bibitem{ma2021contrastive}
S.~Ma, Z.~Zeng, D.~McDuff, and Y.~Song, ``Contrastive learning of global and
  local video representations,'' \emph{NeurIPS}, vol.~34, 2021.

\bibitem{chen2021distilling}
Y.~Chen, Y.~Xian, A.~Koepke, Y.~Shan, and Z.~Akata, ``Distilling audio-visual
  knowledge by compositional contrastive learning,'' in \emph{CVPR}, 2021, pp.
  7016--7025.

\bibitem{shukla2020learning}
A.~Shukla, S.~Petridis, and M.~Pantic, ``Learning speech representations from
  raw audio by joint audiovisual self-supervision,'' \emph{arXiv preprint
  arXiv:2007.04134}, 2020.

\bibitem{zbontar2021barlow}
J.~Zbontar, L.~Jing, I.~Misra, Y.~LeCun, and S.~Deny, ``Barlow twins:
  Self-supervised learning via redundancy reduction,'' in \emph{ICML}.\hskip
  1em plus 0.5em minus 0.4em\relax PMLR, 2021, pp. 12\,310--12\,320.

\bibitem{falcon2022feature}
A.~Falcon, G.~Serra, and O.~Lanz, ``A feature-space multimodal data
  augmentation technique for text-video retrieval,'' in \emph{Proceedings of
  the 30th ACM International Conference on Multimedia}, 2022, pp. 4385--4394.

\bibitem{wang2021dualgnn}
Q.~Wang, Y.~Wei, J.~Yin, J.~Wu, X.~Song, L.~Nie, and M.~Zhang, ``Dualgnn: Dual
  graph neural network for multimedia recommendation,'' \emph{IEEE Transactions
  on Multimedia}, 2021.

\bibitem{qian2021dual}
S.~Qian, D.~Xue, H.~Zhang, Q.~Fang, and C.~Xu, ``Dual adversarial graph neural
  networks for multi-label cross-modal retrieval,'' in \emph{Thirty-Fifth AAAI
  Conference on Artificial Intelligence, AAAI}, 2021, pp. 2440--2448.

\bibitem{saqur2020multimodal}
R.~Saqur and K.~Narasimhan, ``Multimodal graph networks for compositional
  generalization in visual question answering,'' \emph{NeurIPS}, vol.~33, pp.
  3070--3081, 2020.

\bibitem{yang2021multimodal}
X.~Yang, S.~Feng, Y.~Zhang, and D.~Wang, ``Multimodal sentiment detection based
  on multi-channel graph neural networks,'' in \emph{IJCNLP}, 2021, pp.
  328--339.

\bibitem{shirian2022visually}
A.~Shirian, K.~Somandepalli, V.~Sanchez, and T.~Guha, ``Visually-aware acoustic
  event detection using heterogeneous graphs,'' \emph{Proc. Interspeech}, 2022.

\bibitem{wang2021wasserstein}
Y.~Wang, T.~Zhang, X.~Zhang, Z.~Cui, Y.~Huang, P.~Shen, S.~Li, and J.~Yang,
  ``Wasserstein coupled graph learning for cross-modal retrieval,'' in
  \emph{ICCV}, 2021, pp. 1793--1802.

\bibitem{song2021spatial}
X.~Song, J.~Chen, Z.~Wu, and Y.-G. Jiang, ``Spatial-temporal graphs for
  cross-modal text2video retrieval,'' \emph{IEEE Transactions on Multimedia},
  2021.

\bibitem{shirian2021compact}
A.~Shirian and T.~Guha, ``Compact graph architecture for speech emotion
  recognition,'' in \emph{ICASSP}.\hskip 1em plus 0.5em minus 0.4em\relax IEEE,
  2021, pp. 6284--6288.

\bibitem{fu2020magnn}
X.~Fu, J.~Zhang, Z.~Meng, and I.~King, ``Magnn: Metapath aggregated graph
  neural network for heterogeneous graph embedding,'' in \emph{Proceedings of
  The Web Conference}, 2020, pp. 2331--2341.

\bibitem{hu2020heterogeneous}
Z.~Hu, Y.~Dong, K.~Wang, and Y.~Sun, ``Heterogeneous graph transformer,'' in
  \emph{Proceedings of The Web Conference 2020}, 2020, pp. 2704--2710.

\bibitem{lv2021we}
Q.~Lv, M.~Ding, Q.~Liu, Y.~Chen, W.~Feng, S.~He, C.~Zhou, J.~Jiang, Y.~Dong,
  and J.~Tang, ``Are we really making much progress? revisiting, benchmarking
  and refining heterogeneous graph neural networks,'' in \emph{ACM SIGKDD},
  2021, pp. 1150--1160.

\bibitem{kipf2017semi}
T.~N. Kipf and M.~Welling, ``Semi-supervised classification with graph
  convolutional networks,'' in \emph{ICLR}, 2017.

\bibitem{hamilton2017inductive}
W.~Hamilton, Z.~Ying, and J.~Leskovec, ``Inductive representation learning on
  large graphs,'' \emph{Advances in neural information processing systems},
  vol.~30, 2017.

\bibitem{velivckovic2017graph}
P.~Veli{\v{c}}kovi{\'c}, G.~Cucurull, A.~Casanova, A.~Romero, P.~Li{\`o}, and
  Y.~Bengio, ``Graph attention networks,'' in \emph{ICLR}, 2018.

\bibitem{shirian2021dynamic}
A.~Shirian, S.~Tripathi, and T.~Guha, ``Dynamic emotion modeling with learnable
  graphs and graph inception network,'' \emph{IEEE Transactions on Multimedia},
  2021.

\bibitem{gemmeke2017audio}
J.~F. Gemmeke, D.~P. Ellis, D.~Freedman, A.~Jansen, W.~Lawrence, R.~C. Moore,
  M.~Plakal, and M.~Ritter, ``Audio set: An ontology and human-labeled dataset
  for audio events,'' in \emph{ICASSP}, 2017, pp. 776--780.

\bibitem{baevski2020wav2vec}
A.~Baevski, Y.~Zhou, A.~Mohamed, and M.~Auli, ``wav2vec 2.0: A framework for
  self-supervised learning of speech representations,'' \emph{Advances in
  Neural Information Processing Systems}, vol.~33, pp. 12\,449--12\,460, 2020.

\bibitem{tran2018closer}
D.~Tran, H.~Wang, L.~Torresani, J.~Ray, Y.~LeCun, and M.~Paluri, ``A closer
  look at spatiotemporal convolutions for action recognition,'' in \emph{CVPR},
  2018, pp. 6450--6459.

\bibitem{dai2017very}
W.~Dai, C.~Dai, S.~Qu, J.~Li, and S.~Das, ``Very deep convolutional neural
  networks for raw waveforms,'' in \emph{ICASSP}.\hskip 1em plus 0.5em minus
  0.4em\relax IEEE, 2017, pp. 421--425.

\bibitem{akbari2021vatt}
H.~Akbari, L.~Yuan, R.~Qian, W.-H. Chuang, S.-F. Chang, Y.~Cui, and B.~Gong,
  ``Vatt: Transformers for multimodal self-supervised learning from raw video,
  audio and text,'' \emph{arXiv preprint arXiv:2104.11178}, 2021.

\bibitem{shirian2022self}
A.~Shirian, K.~Somandepalli, and T.~Guha, ``Self-supervised graphs for audio
  representation learning with limited labeled data,'' \emph{IEEE Journal of
  Selected Topics in Signal Processing}, vol.~16, no.~6, pp. 1391--1401, 2022.

\bibitem{kong2020panns}
Q.~Kong, Y.~Cao, T.~Iqbal, Y.~Wang, W.~Wang, and M.~D. Plumbley, ``Panns:
  Large-scale pretrained audio neural networks for audio pattern recognition,''
  \emph{IEEE/ACM Transactions on Audio, Speech, and Language Processing},
  vol.~28, pp. 2880--2894, 2020.

\bibitem{gong2021ast}
Y.~Gong, Y.~Chung, and J.~R. Glass, ``{AST:} audio spectrogram transformer,''
  in \emph{Interspeech 2021}.\hskip 1em plus 0.5em minus 0.4em\relax {ISCA},
  2021, pp. 571--575.

\bibitem{simonyan2014very}
K.~Simonyan and A.~Zisserman, ``Very deep convolutional networks for
  large-scale image recognition,'' in \emph{{ICLR}}, Y.~Bengio and Y.~LeCun,
  Eds., 2015.

\end{thebibliography}

\end{document}